\documentclass[aps,prd,reprint,showpacs,nofootinbib,showkeys,superscriptaddress,floatfix]{revtex4-1}

\pdfoutput=1  
\usepackage{soul}
\usepackage{amsmath}
\usepackage{amssymb}
\usepackage{multirow}
\usepackage[american]{babel}
\usepackage{booktabs}
\usepackage{dcolumn}  
\usepackage[T1]{fontenc}  
\usepackage{graphicx}  
\usepackage[]{hyperref}  
\usepackage[utf8]{inputenc}  
\usepackage{url}
\usepackage[dvipsnames]{xcolor}
\usepackage{float}
\usepackage[normalem]{ulem}

\DeclareUnicodeCharacter{2212}{-}

\usepackage{tikz}

\usetikzlibrary{arrows,shapes,positioning,shadows,trees}

\tikzset{
	basic/.style  = {draw, text width=2cm, drop shadow, font=\sffamily, rectangle},
	root/.style   = {basic, rounded corners=2pt, thin, align=center,
		fill=green!30},
	level 2/.style = {basic, rounded corners=6pt, thin,align=center, fill=green!60,
		text width=8em},
	level 3/.style = {basic, thin, align=left, fill=pink!60, text width=6.5em}
}

\usetikzlibrary{arrows,decorations.pathmorphing,backgrounds,fit,positioning,shapes.symbols,chains}

\newcolumntype{d}[1]{D{.}{.}{#1}}
\newcolumntype{v}[1]{D{,}{,\ }{#1}}

\makeatletter

\newcommand{\Rmnum}[1]{\expandafter\@slowromancap\romannumeral #1@}

\makeatother

\renewcommand {\arraystretch}{1.3}

\usepackage{hyperref}
\hypersetup{
    colorlinks=true, 
    linkcolor=blue, 
    citecolor=magenta}

\begin{document}

\title{Interacting dark energy after DESI DR2: a challenge for the $\Lambda$CDM paradigm? }

\author{Supriya Pan}
\email{supriya.maths@presiuniv.ac.in}
\affiliation{Department of Mathematics, Presidency University, 86/1 College Street, Kolkata 700073, India}
\affiliation{Institute of Systems Science, 
Durban University of Technology, Durban 4000, Republic of South Africa}

\author{Sivasish Paul}
\email{sivasishpaul@gmail.com}
\affiliation{Department of Mathematics, Presidency University, 86/1 College Street,  Kolkata 700073, India}

\author{Emmanuel N. Saridakis}
\email{msaridak@noa.gr}
\affiliation{National Observatory of Athens, Lofos Nymfon, 11852 Athens, Greece}
\affiliation{Departamento de Matem\'{a}ticas, Universidad Cat\'{o}lica del 
Norte, Avda.
Angamos 0610, Casilla 1280 Antofagasta, Chile}
\affiliation{CAS Key Laboratory for Researches in Galaxies and Cosmology, 
Department of
Astronomy, University of Science and Technology of China, Hefei, Anhui
230026, P.R. China}

\author{Weiqiang Yang}
\email{d11102004@163.com}
\affiliation{Department of Physics, Liaoning Normal University, Dalian, 116029, 
P. R.  China}


\begin{abstract}

We investigate the scenario of interacting dark energy through a detailed 
confrontation with various observational datasets. We quantify the interaction 
in a general way, through the deviation from the standard  scaling 
of the dark matter energy density. We use the cosmic microwave background (CMB) data from Planck 2018, data from Baryon Acoustic Oscillations 
(BAO) from the recently released DESI DR2,
 observational Hubble Data 
from Cosmic Chronometers (CC),  
and finally various Supernova Type Ia (SNIa) datasets (PantheonPlus, Union3 and DESY5). For the basic and simplest interacting model we find a mild preference of the interaction at slightly more than $1\sigma$ however still within $2\sigma$, and thus no strong evidence of interaction is found.   However, comparison  with $\Lambda$ cold dark matter ($\Lambda$CDM) scenario through $\Delta \chi^2_{\rm min}$,  Akaike Information Criterion  and Bayesian analysis, reveals a mixed picture, namely  according to $\Delta \chi^2_{\rm min}$ the interaction is mildly favored by most of the datasets, while the remaining statistical measures are inclined toward $\Lambda$CDM. 

\end{abstract}
\maketitle

\section{Introduction}
\label{sec1}

According to different accumulating  observations 
our universe has experienced 
two phases of  accelerating expansion, one at the primordial phase and one at 
late times. The late-time acceleration can indeed be explained by the 
simple cosmological constant, however the relevant theoretical problem, the 
possibility of a dynamical nature, as well as various potential observational  
tensions    
\cite{DiValentino:2021izs,Perivolaropoulos:2021jda,Schoneberg:2021qvd,Abdalla:2022yfr,Kamionkowski:2022pkx}, led to consider scenarios that 
go beyond $\Lambda$-Cold Dark Matter ($\Lambda$CDM) paradigm or/and beyond general relativity. A first 
avenue that one can follow is  to 
construct  modified gravitational theories, which have 
general relativity as a limiting case, but which in general can alter the 
universe evolution by providing new  additional degree(s) of freedom   
\cite{Capozziello:2011et,CANTATA:2021asi}. Another way is to maintain general 
relativity but introduce new  sectors such as the  dark 
energy one or other exotic forms of matter 
\cite{Copeland:2006wr,Cai:2009zp}, which can give rise to 
acceleration.

Since the underlying microscopical theory of dark energy is unknown,  and since 
the underlying microscopical theory of dark matter is unknown too,  there is no 
reason that could theoretically forbid their mutual interaction. From the 
particle physics and field theoretical point of view, the investigation of such 
 interactions could be enlightening for revealing the nature of the dark 
sectors. Nevertheless,  from the  cosmological point of view, the  
possibility of an interacting dark energy has the additional advantage that it 
can alleviate the problems of $\Lambda$CDM paradigm, such as the coincidence one
\cite{Amendola:1999qq,Amendola:2000uh,Billyard:2000bh,
Zimdahl:2001ar,Wetterich:1994bg,Amendola:1999er, 
Cai:2004dk, delCampo:2008sr}, the presence of
accelerating scaling solution  
\cite{Chimento:2003iea,Barrow:2006hia,Amendola:2006dg,delCampo:2008jx,
He:2008tn,Chen:2008ft, Basilakos:2008ae,
Gavela:2009cy,
Jamil:2009eb,Lip:2010dr,Chen:2011cy, Yang:2014gza, 
Faraoni:2014vra,Salvatelli:2014zta,Yang:2014hea,Pan:2012ki, Nunes:2016dlj,
Yang:2016evp, Mukherjee:2016shl,Cai:2017yww,Yang:2017zjs,Santos:2017bqm,Pan:2017ent,Xu:2017rfo,Yang:2018xlt,vonMarttens:2018iav, Pan:2019jqh, 
Li:2019loh, Pan:2019gop,
Yang:2019bpr,Yang:2019vni,Khyllep:2021wjd,Halder:2024aan}, the   $H_0$ tension
  \cite{DiValentino:2015ola,Kumar:2017dnp,
Khosravi:2017hfi,Mortsell:2018mfj,  
Guo:2018ans,
Yang:2018qmz,
 Kreisch:2019yzn,Martinelli:2019dau,Pandey:2019plg,
Vattis:2019efj,Agrawal:2019lmo, 
Yang:2019nhz,Giare:2024smz}, and the $\sigma_8$  tension
\cite{Pourtsidou:2016ico,An:2017crg,DiValentino:2018gcu,Kumar:2019wfs,Khoury:2025txd}  
(for reviews see 
\cite{Bolotin:2013jpa,Wang:2016lxa}).

The smoking gun of interacting dark energy is an altered evolution law of the dark matter energy density $\rho_{\rm dm}$, since due to the interaction the 
latter is no more  conserved.
In particular, considering dark matter as a cold barotropic fluid,  one 
has 
\begin{equation}\label{DM-evolution}
\rho_{\rm dm} = \rho_{\rm dm,0}\; a^{-3+\delta(a)},
\end{equation}
 with $\rho_{\rm dm,0}$   the dark
matter energy density at present value, $a$ the scale factor, and where the 
deviation from $\Lambda$CDM paradigm, namely from the non-interacting case, is 
quantified by $\delta(a)$, which in general can be a varying quantity.   If $\delta (a) < 0$, then for increasing  scale factor  $\rho_{\rm dm}$ decreases more rapidly than its standard evolution $a^{-3}$, which implies an energy flow from dark matter to dark energy. On the other hand, for $\delta (a) > 0$ we have the opposite situation.

In the present letter, we desire to confront interacting dark energy with observational data, including the recent Data Release 2 (DR2)  of Dark Energy Spectroscopic Instrument (DESI) ~\cite{Lodha:2025qbg}.\footnote{From Data Release 1~\cite{DESI:2024mwx} to Data Release 2~\cite{DESI:2025zgx,Lodha:2025qbg}, DESI has indicated the preference for a dynamical DE 
(assuming the Chevallier-Polarski-Linder parametrization~\cite{Chevallier:2000qy, Linder:2002et}) and prompted widespread discussion 
in the cosmology community~\cite{Yin:2021uus, Cortes:2024lgw,Luongo:2024fww,Gialamas:2024lyw,Colgain:2024xqj,Tada:2024znt,Carloni:2024zpl,Alfano:2024fzv,Notari:2024rti,Giare:2024gpk,Reboucas:2024smm,Jiang:2024xnu,Li:2024qus,Notari:2024zmi,Colgain:2024mtg, Yin:2024hba, Wolf:2025jlc,Huang:2025som,Ormondroyd:2025iaf,Brandenberger:2025hof,Odintsov:2025kyw,Luu:2025fgw,Nakagawa:2025ejs,Pan:2025psn,Paliathanasis:2025dcr,Pang:2025lvh,Nesseris:2025lke,Chaussidon:2025npr,Wang:2025ljj,Yang:2025mws, Alfano:2025gie,You:2025uon, Lee:2025yvn}. 
} As we show, we find that a non-zero interaction is favored, which reveals a potential challenge for $\Lambda$CDM paradigm. The plan of the work is the following. In Section \ref{sec2}, we present the scenario of interacting dark energy. Then, in Section \ref{sec-data+results} we provide the observational datasets that we use in our analysis, and we perform the full observational confrontation, deriving the corresponding contour plots. 
Finally, in Section \ref{sec-summary} we discuss our results and we conclude.

\section{Interacting dark energy}
\label{sec2}

Let us present the scenario of interacting dark energy in the framework of 
general relativity. We focus on the    spatially flat 
Friedmann-Robertson-Walker (FRW) metric, described by the line element, $ds^2 = 
-dt^2 + a^2 (t) (dx^2 + dy^2 + dz^2)$, where $a$ is the scale factor.
The general Friedmann equations can be written as 

\begin{eqnarray}
&&H^2  = \frac{8\pi G}{3}  \left( \rho_{\rm b} + \rho_{\rm r} + \rho_{\rm \nu} +  \rho_{\rm dm} + \rho_{\rm de} 
\right),\\
&& 2 \dot{H}+3H^2 = - 8 \pi G 
 (p_{\rm b} + p_{\rm r} + p_{\rm \nu} + p_{\rm dm} +p_{\rm de}),
\end{eqnarray}
where $H=\dot{a}/a$ is the Hubble function (an overhead dot represents the cosmic time differentiation). The quantities $\rho_{i}$ and $p_i$  represent the energy density and pressure of the $i$-th fluid component, respectively, with $i= {\rm b}, {\rm r}, {\rm \nu}, {\rm dm}, {\rm de}$ corresponding to baryons (b), radiation (r), neutrinos ($\nu$) (assuming one massive neutrino with a fixed mass of 0.06 eV and two massless neutrinos), cold dark matter (dm)  and dark energy (de). Since dark matter is assumed to be cold, we set $p_{\rm dm} = 0$.

In the case where an interaction between dark energy (DE) and dark matter  (DM) sectors is present, due to the diffeomorphism invariant the total energy-momentum tensor is indeed conserved, however, the separate DE and DM sectors do not.
In particular, in such an interacting  case one has 
 \begin{align}\label{coupled-eqns}
\dot{\rho}_{\rm dm} + 3 H \rho_{\rm dm} =  - 
\dot{\rho}_{\rm de} - 3 H (1+w_{\rm de}) \rho_{\rm de} = - Q,
\end{align}
with $w_{\rm de} = p_{\rm de}/\rho_{\rm de}$ being the barotropic equation-of-state 
parameters of the DE sector, 
and where  $Q$ quantifies    the 
interaction rate (also known as the interaction function) between the dark 
sectors. 
Usually, there is no 
guiding principle to determine  $Q$, and thus in the literature one may find  
 several phenomenological choices 
\cite{Chimento:2003iea, Shapiro:2003ui, Sola:2005hb, Barrow:2006hia,Amendola:2006dg,delCampo:2008jx,He:2008tn,Chen:2008ft, Basilakos:2008ae,
Gavela:2009cy,Jamil:2009eb,Lip:2010dr,Chen:2011cy, Yang:2014gza,Faraoni:2014vra,Salvatelli:2014zta,Yang:2014hea,Shahalam:2015sja,Pan:2012ki,Gomez-Valent:2014rxa, Sola:2015wwa, Sola:2016jky, SolaPeracaula:2016qlq,  Nunes:2016dlj,Yang:2016evp, Mukherjee:2016shl,Cai:2017yww,Yang:2017zjs,Santos:2017bqm,
Pan:2017ent,Xu:2017rfo, SolaPeracaula:2017esw, SolaPeracaula:2017esw, Zhang:2018glx,Yang:2018xlt,vonMarttens:2018iav,Pan:2019jqh,Li:2019loh,Yang:2019bpr,Yang:2019vni,Liu:2022hpz,Zhao:2022ycr, SolaPeracaula:2022hpd, Li:2024qso}.
 However, as we mentioned in the Introduction, in full generality one may 
quantify the effect of the interaction straightaway in the scaling behavior, 
which for cold dark matter, 
will take the form 
(\ref{DM-evolution}).

In this work we focus on the basic interacting scenario in which   $w_{\rm de} 
=-1$, namely DE corresponds to the vacuum energy sector, and  $\delta(a) = 
\delta_0=const$. Inserting (\ref{DM-evolution}) with $\delta(a) = 
\delta_0$  into the conservation equation (\ref{coupled-eqns}) for DM, we 
extract an  interaction function of the form
\begin{eqnarray}
Q = -\delta_0  H \rho_{\rm dm} .
\end{eqnarray}
 Additionally,  for the energy density of DE we have 
\begin{eqnarray}
\rho_{\rm de} (a) = \rho_{\rm de,0} + \rho_{\rm dm, 0} \left(\frac{\delta_0 }{-3 
+\delta_0}\right)  \times \left[ 1 - a^{-3 + \delta_0} \right].
\end{eqnarray}
Note that for $\delta_0 =3$  the energy density for DE is undefined, while we have a constant DM energy density, case which is not physically interesting, and thus in the following   we exclude  $\delta_0 =3$ from our parameter space. Now, introducing the density parameters $\Omega_{i}=8 \pi G \rho_{i}/(3H^2)$ (where $i$ stands for baryons,  radiation, neutrinos, DM and DE), we can write the first Friedmann equation as

   \begin{eqnarray}\label{Hubble-expansion}
\left(\frac{H}{H_0} \right)^2 = \Omega_{b0}a^{-3}+ \Omega_{r0}a^{-4}+\Omega_{\nu, 0}\frac{\rho_{\nu}}{\rho_{\nu,0}} +  \Omega_{\rm dm,0} a^{-3 +\delta_0} \nonumber\\+ \Omega_{\rm 
de,0} + \frac{\delta_0 \Omega_{\rm dm,0}}{-3+\delta_0}  \left( 1- a^{-3 + 
\delta_0} \right),
\end{eqnarray}
where $\Omega_{i,0}$ denotes the present-day value of $\Omega_i$.
Notice that the Hubble expansion law (\ref{Hubble-expansion}) can be interpreted  as the solution of two non-interacting cosmic fluids as follows: the conservation equations in (\ref{coupled-eqns}) can be rewritten as $\dot{\rho}_{\rm dm} + 3 H (1+w_{\rm dm}^{\rm eff}) \rho_{\rm dm} =0$, and $\dot{\rho}_{\rm de} + 3 H (1+w_{\rm de}^{\rm eff}) \rho_{\rm de} =0$, where $w_{\rm dm}^{\rm eff} \equiv Q/(3H \rho_{\rm dm})$ and $w_{\rm de}^{\rm eff} = -1 - Q/(3H\rho_{\rm de})$ are respectively the effective equation-of-state parameters of DM and DE. For the present set-up, $w_{\rm dm}^{\rm eff} = -\delta_0/3$ and $w_{\rm de}^{\rm eff} = -1 + \frac{\delta_0}{3} \times \frac{\rho_{\rm dm}}{\rho_{\rm de}} =  (\delta_0/3) \times a^{-3 +\delta_0} \left[(\Omega_{\rm de,0}/\Omega_{\rm dm,0}) + (\delta_0/(-3+\delta_0)) \times (1-a^{-3+\delta_0}) \right]^{-1}$.  This emphasizes that in presence of an interaction, DM may not remain cold effectively.

Having established the background evolution of the interacting dark energy (IDE) framework, we now focus on the study of linear perturbations. To analyze the evolution of inhomogeneities, we employ the synchronous gauge, where scalar perturbations are introduced into the spatially flat Friedmann–Robertson–Walker (FRW) metric, expressed as ~\cite{Ma:1995ey, Yang:2020zuk}: 
\begin{eqnarray}
ds^2 &=& a^2(\tau) \Big[ 
- d\tau^2 + (\delta_{ij} + h_{ij})dx^{i} dx^{j} \, \Big],
\end{eqnarray}
where $\tau$ represents the conformal time, $a(\tau)$ is the scale factor in conformal time, and $\delta_{ij}$ and $h_{ij}$ correspond to the background and perturbed components of the spatial metric, respectively.

In the present context, since dark energy corresponds to the vacuum energy  it does not contribute to the perturbations. Consequently, both the DE density perturbation and its velocity divergence vanish ($\delta_{\rm de} = 0$, $\theta_{\rm de} = 0$), indicating that the evolution of perturbations is governed solely by the cold DM sector. 
Following the formulations in Refs.~\cite{Wang:2014xca, Wang:2018azy, Pan:2019jqh}, the corresponding first-order perturbation equations can be expressed as,
	\begin{eqnarray}
		&&\delta _{\rm dm}^{\prime } =-\left( \theta _{\rm dm}+\frac{h^{\prime }}{2}%
		\right)+\delta_0\left(\theta+\frac{h^\prime}{2}\right), \\
		&&\theta _{\rm dm}^{\prime } =-\mathcal{H}\theta_{\rm dm},
	\end{eqnarray}
where a prime denotes the derivative with respect to the conformal time $\tau$. 
Here, $\delta_{\rm dm}$ and $\theta_{\rm dm}$ are the density contrast and the velocity divergence of CDM, respectively, $h$ is the trace of metric perturbation $h_{ij}$, 
$\mathcal{H} = a^{\prime} (\tau)/a (\tau)$ is the conformal Hubble parameter, and $\theta$ represents the divergence of the peculiar velocity field in Fourier space.
The additional term proportional to $\delta_0$ accounts for the contribution arising from the interaction term.  At the linear perturbation level, following  Ref.~\cite{Wands:2012vg, De-Santiago:2012xpd, Wang:2013qy, Wang:2014xca, Yang:2020zuk}, we assume that the energy transfer four-vector is parallel to cold DM four-velocity, i.e. $Q^{\mu} \propto u^{\mu}$, which corresponds to energy exchange without any momentum transfer. This assumption ensures that dark matter follows geodesic motion in its own co-moving frame, thereby eliminating any force term acting on dark matter particles. As a consequence, the vacuum energy remains homogeneous, leading to vanishing vacuum perturbations. To fix the residual gauge freedom inherent to the synchronous gauge, we consider the CDM frame, setting $\theta_{\rm dm} = 0$, and consequently $\theta^{\prime}_{\rm dm} = 0$.

\begin{table}
	\begin{center}
		\renewcommand{\arraystretch}{1.4}
		\begin{tabular}{|c@{\hspace{1 cm}}|@{\hspace{1 cm}} c|}
			\hline
			\textbf{Parameter}           & \textbf{Prior}\\
			\hline\hline
			
			$\Omega_{\rm dm} h^2$            & $[0.01,0.99]$ \\

            $\Omega_{b} h^2$             & $[0.005,0.1]$ \\

			$\tau$                       & $[0.01,0.8]$ \\
			$n_s$                        & $[0.8, 1.2]$ \\
			$\log[10^{10}A_{s}]$         & $[1.61,3.91]$ \\
			$100\theta_{MC}$             & $[0.5,10]$ \\ 
			$\delta_0$                        & $[-1, 1]$ \\
			
			\hline
		\end{tabular}
	\end{center}
	\caption{We show the prior used on the cosmological parameters varied independently during the statistical analysis. }
	\label{tab:priors}
\end{table}

\begingroup
\begin{center}
	\begin{table*}
		\scalebox{0.85}{
			\begin{tabular}{ccccccccc}
				\hline\hline
				Parameters & CMB & CMB+DESI & CMB+DESI+PantheonPlus & CMB+DESI+Union3 & CMB+DESI+DESY5\\ \hline

				$\Omega_{\rm dm} h^2$ & 
				$0.11743_{-0.00066-0.00132}^{+0.00068+0.00132}$ & 
				$0.11770_{-0.00066-0.00130}^{+0.00066+0.00130}$ & 
				$0.11722_{-0.00066-0.00131}^{+0.00066+0.00128}$ & 
				$0.11751_{-0.00064-0.00127}^{+0.00065+0.00127}$ & 
				$0.11747_{-0.00066-0.00127}^{+0.00067+0.00128}$ \\

                $\Omega_b h^2$ & 
				$0.02241_{-0.00017-0.00032}^{+0.00017+0.00033}$ & 
				$0.02235_{-0.00016-0.00032}^{+0.00016+0.00031}$ & 
				$0.02236_{-0.00016-0.00032}^{+0.00016+0.00032}$ & 
				$0.02236_{-0.00017-0.00031}^{+0.00016+0.00033}$ & 
				$0.02236_{-0.00016-0.00031}^{+0.00016+0.00032}$ \\
				
				$100\theta_{MC}$ & 
				$1.04025_{-0.00037-0.00079}^{+0.00041+0.00073}$ & 
				$1.04092_{-0.00028-0.00054}^{+0.00028+0.00054}$ & 
				$1.04091_{-0.00028-0.00055}^{+0.00028+0.00055}$ & 
				$1.04092_{-0.00029-0.00057}^{+0.00028+0.00055}$ & 
				$1.04090_{-0.00028-0.00055}^{+0.00028+0.00056}$ \\
				
				$\tau$ & 
				$0.0530_{-0.0077-0.0146}^{+0.0072+0.0154}$ & 
				$0.0571_{-0.0084-0.0151}^{+0.0072+0.0168}$ & 
				$0.0573_{-0.0078-0.0154}^{+0.0079+0.0162}$ & 
				$0.0569_{-0.0080-0.0152}^{+0.0079+0.0167}$ & 
				$0.0569_{-0.0082-0.0155}^{+0.0073+0.0166}$ \\
				
				$n_s$ & 
				$0.9674_{-0.0050-0.0100}^{+0.0051+0.0099}$ & 
				$0.9774_{-0.0035-0.0070}^{+0.0036+0.0071}$ & 
				$0.9772_{-0.0036-0.0069}^{+0.0036+0.0072}$ & 
				$0.9773_{-0.0034-0.0070}^{+0.0035+0.0069}$ & 
				$0.9769_{-0.0034-0.0068}^{+0.0035+0.0067}$ \\
				
				${\rm{ln}}(10^{10} A_s)$ & 
				$3.053_{-0.015-0.030}^{+0.015+0.032}$ & 
				$3.057_{-0.017-0.031}^{+0.015+0.034}$ & 
				$3.058_{-0.016-0.032}^{+0.016+0.033}$ & 
				$3.057_{-0.016-0.033}^{+0.016+0.034}$ & 
				$3.057_{-0.017-0.032}^{+0.015+0.034}$ \\
				
				$\delta_0$ & 
				$0.0013_{-0.0008-0.0016}^{+0.0008+0.0015}$ & 
				$-0.00059_{-0.00034-0.00067}^{+0.00034+0.00068}$ & 
				$-0.00050_{-0.00034-0.00066}^{+0.00033+0.00067}$ & 
				$-0.00051_{-0.00034-0.00066}^{+0.00034+0.00066}$ & 
				$-0.00044_{-0.00033-0.00066}^{+0.00034+0.00066}$ \\
				
				$\Omega_{m0}$ & 
				$0.3623_{-0.0302-0.0523}^{+0.0256+0.0589}$ & 
				$0.2968_{-0.0042-0.0085}^{+0.0043+0.0089}$ & 
				$0.2988_{-0.0044-0.0086}^{+0.0044+0.0087}$ & 
				$0.2985_{-0.0044-0.0086}^{+0.0045+0.0088}$ & 
				$0.3006_{-0.0044-0.0086}^{+0.0044+0.0089}$ \\
				
				$\sigma_8$ & 
				$0.789_{-0.015-0.030}^{+0.015+0.031}$ & 
				$0.817_{-0.011-0.021}^{+0.011+0.022}$ & 
				$0.816_{-0.011-0.022}^{+0.011+0.022}$ & 
				$0.816_{-0.011-0.022}^{+0.011+0.021}$ & 
				$0.815_{-0.011-0.021}^{+0.011+0.022}$ \\
				
				$H_0$ [Km/s/Mpc] & 
				$64.06_{-1.79-3.51}^{+1.79+3.61}$ & 
				$68.74_{-0.37-0.76}^{+0.37+0.74}$ & 
				$68.56_{-0.37-0.73}^{+0.37+0.74}$ & 
				$68.59_{-0.38-0.75}^{+0.37+0.75}$ & 
				$68.42_{-0.37-0.76}^{+0.37+0.73}$ \\
				
				$S_8$ & 
				$0.866_{-0.024-0.045}^{+0.024+0.047}$ & 
				$0.813_{-0.011-0.020}^{+0.010+0.021}$ & 
				$0.814_{-0.011-0.022}^{+0.011+0.022}$ & 
				$0.814_{-0.011-0.023}^{+0.011+0.022}$ & 
				$0.815_{-0.011-0.021}^{+0.011+0.022}$ \\
				
				$r_{\rm{drag}}$ [Mpc] & 
				$146.72_{-0.36-0.74}^{+0.38+0.72}$ & 
				$147.51_{-0.23-0.46}^{+0.23+0.45}$ & 
				$147.50_{-0.23-0.44}^{+0.22+0.45}$ & 
				$147.49_{-0.23-0.44}^{+0.23+0.44}$ & 
				$147.48_{-0.22-0.43}^{+0.22+0.43}$ \\

				\hline\hline                                   
\end{tabular}}                                                                
		\caption{68\% and 95\% CL constraints on the model parameters of the present interacting scenario considering CMB alone and its inclusion with DESI DR2 and three different compilations of SNIa (PantheonPlus, Union3 and DESY5).}
		\label{tab:IDE-noCC}                   
	\end{table*}                                     
\end{center}
\endgroup

\begingroup
\begin{center}
	\begin{table*}
		\scalebox{0.85}{
			\begin{tabular}{ccccccccc}
				\hline\hline
				Parameters & CMB+CC+DESI & CMB+CC+DESI+PantheonPlus & CMB+CC+DESI+Union3 & CMB+CC+DESI+DESY5\\ \hline

				$\Omega_{\rm dm} h^2$ & 
				$0.11722_{-0.00066-0.00131}^{+0.00066+0.00128}$ & 
				$0.11751_{-0.00064-0.00127}^{+0.00065+0.00127}$ & 
				$0.11747_{-0.00066-0.00127}^{+0.00067+0.00128}$ & 
				$0.11774_{-0.00064-0.00130}^{+0.00065+0.00126}$ \\
                
                $\Omega_b h^2$ & 
				$0.02235_{-0.00016-0.00032}^{+0.00016+0.00031}$ & 
				$0.02235_{-0.00016-0.00032}^{+0.00016+0.00033}$ & 
				$0.02235_{-0.00016-0.00033}^{+0.00016+0.00033}$ & 
				$0.02236_{-0.00016-0.00032}^{+0.00016+0.00032}$ \\
				
				$100\theta_{MC}$ & 
				$1.04093_{-0.00028-0.00053}^{+0.00028+0.00055}$ & 
				$1.04092_{-0.00028-0.00054}^{+0.00028+0.00055}$ & 
				$1.04091_{-0.00028-0.00053}^{+0.00028+0.00055}$ & 
				$1.04091_{-0.00028-0.00055}^{+0.00028+0.00055}$ \\
				
				$\tau$ & 
				$0.0573_{-0.0083-0.0156}^{+0.0075+0.0165}$ & 
				$0.0570_{-0.0080-0.0149}^{+0.0072+0.0162}$ & 
				$0.0571_{-0.0082-0.0154}^{+0.0076+0.0162}$ & 
				$0.0571_{-0.0083-0.0151}^{+0.0074+0.0167}$ \\
				
				$n_s$ & 
				$0.9774_{-0.0035-0.0068}^{+0.0035+0.0069}$ & 
				$0.9771_{-0.0035-0.0069}^{+0.0035+0.0067}$ & 
				$0.9771_{-0.0035-0.0070}^{+0.0035+0.0071}$ & 
				$0.9770_{-0.0035-0.0069}^{+0.0036+0.0068}$ \\
				
				${\rm{ln}}(10^{10} A_s)$ & 
				$3.058_{-0.017-0.032}^{+0.016+0.033}$ & 
				$3.058_{-0.016-0.031}^{+0.016+0.033}$ & 
				$3.058_{-0.016-0.031}^{+0.016+0.033}$ & 
				$3.057_{-0.017-0.032}^{+0.015+0.034}$ \\
				
				$\delta_0$ & 
				$-0.00058_{-0.00033-0.00064}^{+0.00034+0.00066}$ & 
				$-0.00050_{-0.00033-0.00066}^{+0.00034+0.00065}$ & 
				$-0.00051_{-0.00034-0.00067}^{+0.00034+0.00068}$ & 
				$-0.00041_{-0.00033-0.00064}^{+0.00033+0.00066}$ \\
				
				$\Omega_{m0}$ & 
				$0.2968_{-0.0043-0.0086}^{+0.0043+0.0086}$ & 
				$0.2990_{-0.0043-0.0084}^{+0.0043+0.0085}$ & 
				$0.2988_{-0.0044-0.0085}^{+0.0043+0.0089}$ & 
				$0.3010_{-0.0042-0.0086}^{+0.0043+0.0084}$ \\
				
				$\sigma_8$ & 
				$0.817_{-0.011-0.021}^{+0.011+0.022}$ & 
				$0.816_{-0.011-0.021}^{+0.011+0.022}$ & 
				$0.816_{-0.011-0.021}^{+0.011+0.022}$ & 
				$0.814_{-0.011-0.021}^{+0.011+0.021}$ \\
				
				$H_0$ [Km/s/Mpc] & 
				$68.73_{-0.37-0.74}^{+0.37+0.75}$ & 
				$68.55_{-0.37-0.71}^{+0.37+0.73}$ & 
				$68.57_{-0.37-0.75}^{+0.37+0.72}$ & 
				$68.39_{-0.36-0.70}^{+0.36+0.73}$ \\
				
				$S_8$ & 
				$0.813_{-0.011-0.021}^{+0.011+0.021}$ & 
				$0.814_{-0.011-0.021}^{+0.011+0.021}$ & 
				$0.814_{-0.011-0.021}^{+0.011+0.022}$ & 
				$0.815_{-0.011-0.021}^{+0.011+0.021}$ \\
				
				$r_{\rm{drag}}$ [Mpc] & 
				$147.51_{-0.22-0.42}^{+0.22+0.43}$ & 
				$147.48_{-0.22-0.43}^{+0.22+0.44}$ & 
				$147.49_{-0.23-0.44}^{+0.23+0.43}$ & 
				$147.48_{-0.22-0.44}^{+0.22+0.44}$ \\

				\hline\hline                                                         
				
		\end{tabular}}                                                                
		\caption{68\% and 95\% CL constraints on the model parameters of the present interacting scenario for CMB+CC+DESI and CMB+CC+DESI+SNIa, where SNIa is either PantheonPlus or Union3 or DESY5. }
		\label{tab:IDE-withCC}                   
	\end{table*}                                     
\end{center}
\endgroup

\section{Data and results}
\label{sec-data+results}

\begin{figure*}
    \centering
\includegraphics[width=0.85\textwidth]{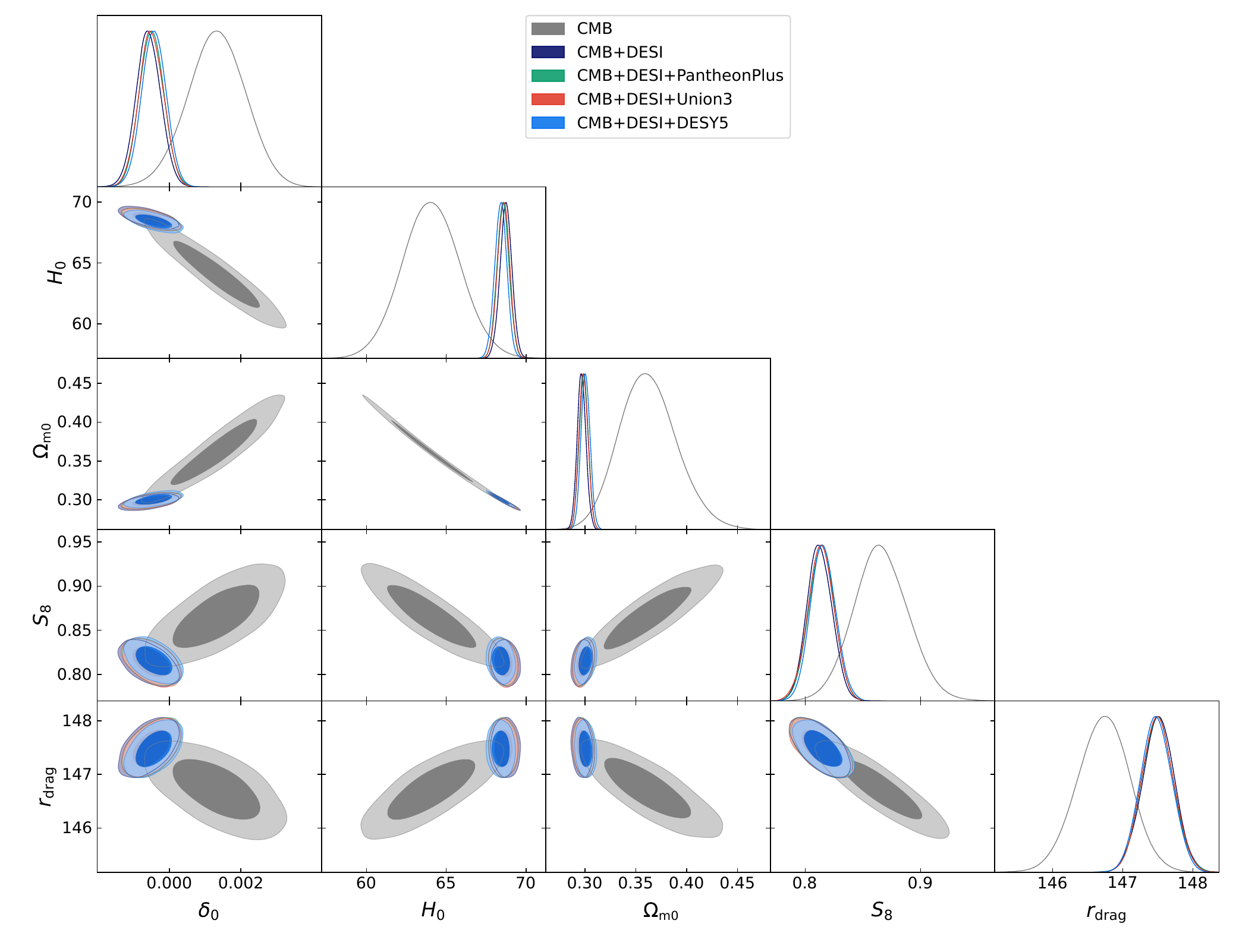}
\caption{One-dimensional marginalized posterior distributions and two-dimensional joint contours of the model parameters considering CMB, CMB+DESI, CMB+DESI+SNIa, where SNIa is either PantheonPlus or Union3 or DESY5.}
\label{fig:contour1}
\end{figure*}
\begin{figure*}
\includegraphics[width=0.85\textwidth]{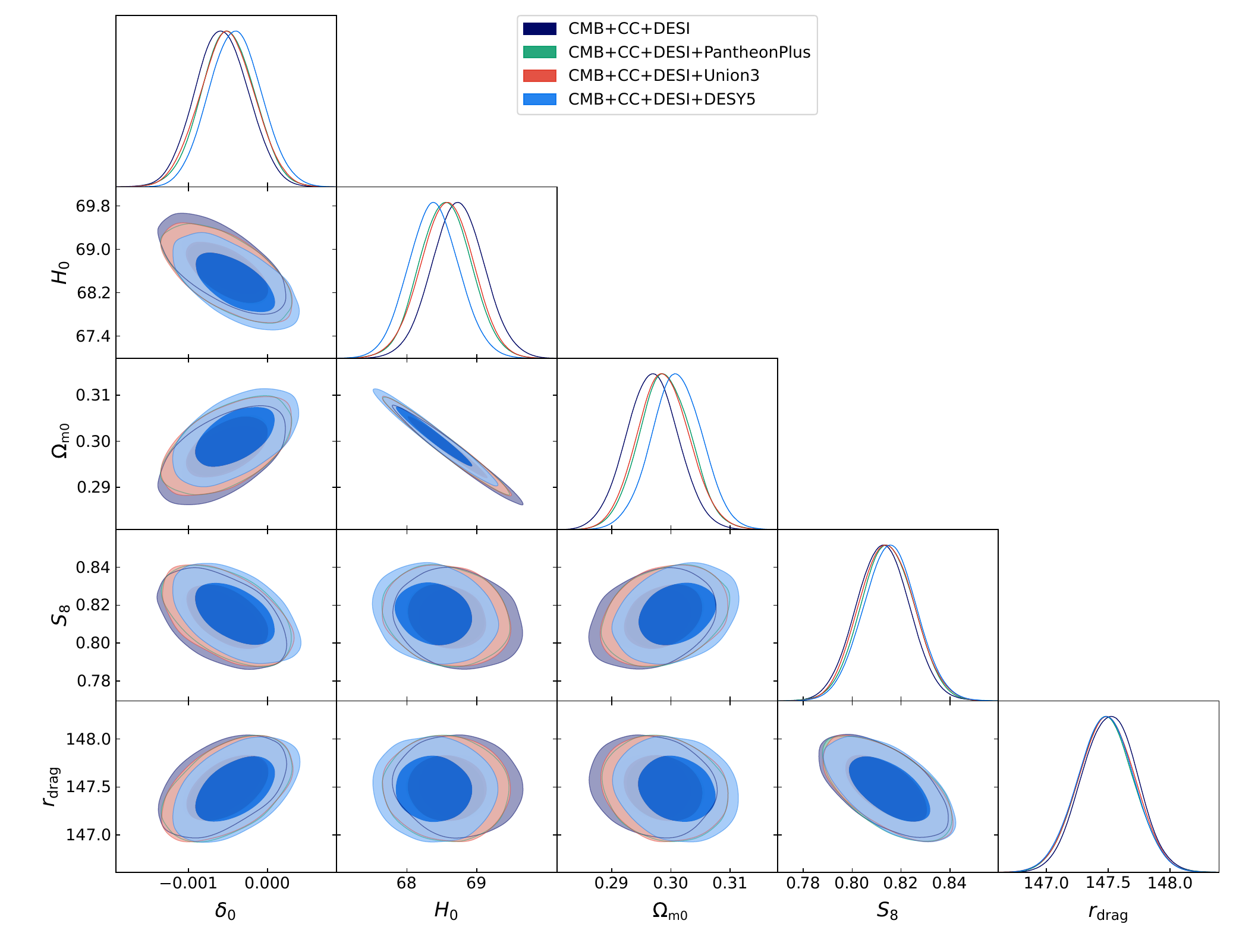}
\caption{One-dimensional marginalized posterior distributions and two-dimensional joint contours of the model parameters considering CMB+CC+DESI and CMB+CC+DESI+SNIa where SNIa is either PantheonPlus or Union3 or DESY5. }  
    \label{fig:contour2}
\end{figure*}
\begin{figure}
    \centering
    \includegraphics[width=0.48\textwidth]{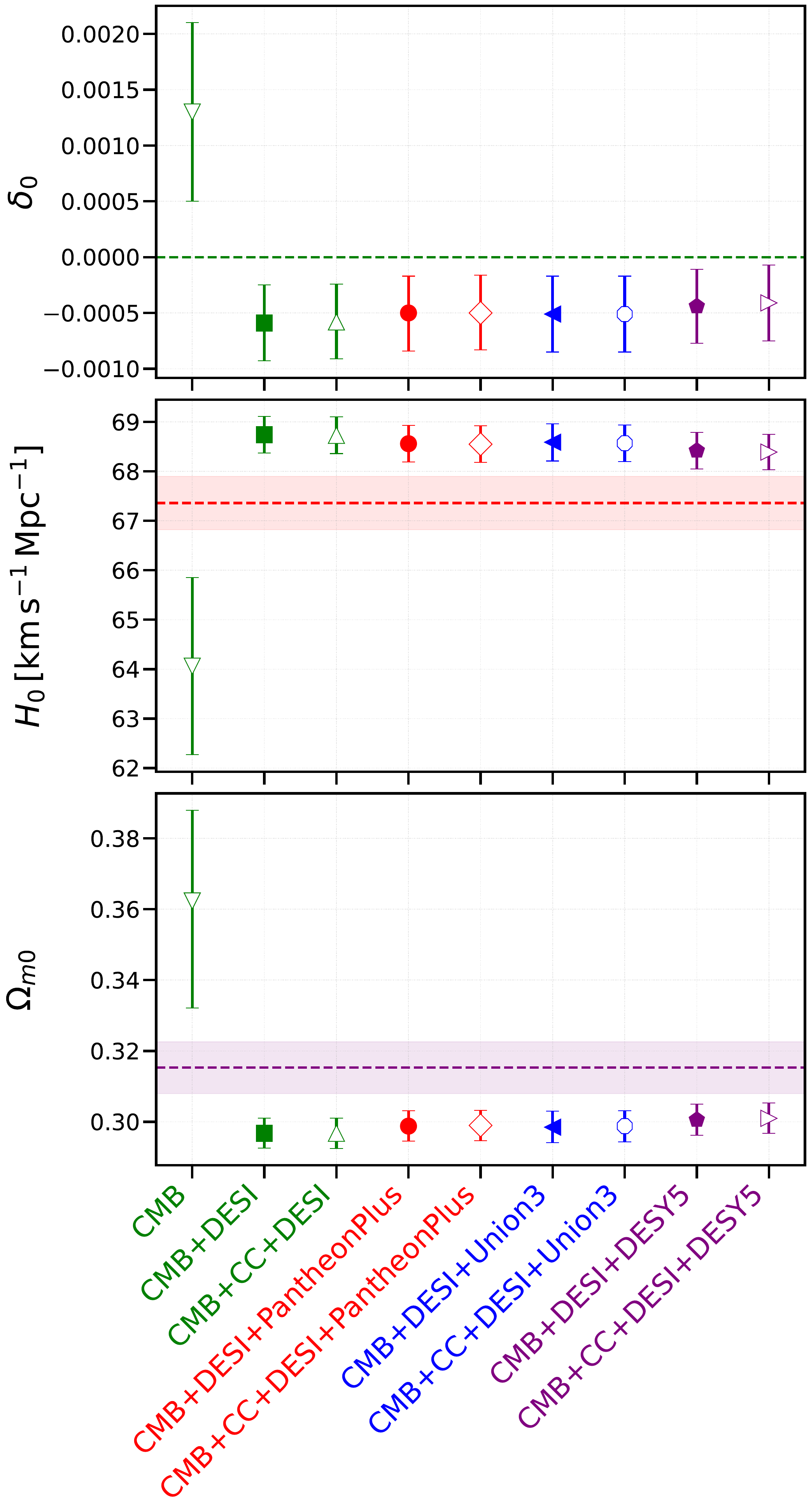}
    \caption{Whisker graphs of $\delta_0$ (upper plot), $H_0$ (middle plot) and $\Omega_{m0}$ (lower plot),  with their  68\% CL constraints, for various datasets used in our analysis. The horizontal bar in the middle panel (red) corresponds to $H_0 = 67.36\, \pm \, 0.54 $ km/s/Mpc at 68\% CL \cite{Planck:2018vyg} and the horizontal bar in the lower plot (violet) corresponds to $\Omega_{m0} = 0.3153\, \pm \, 0.0073$ at 68\% CL~\cite{Planck:2018vyg}.  }
    \label{fig:whisker}
\end{figure}
\begin{figure}
    \includegraphics[width=0.5\textwidth]{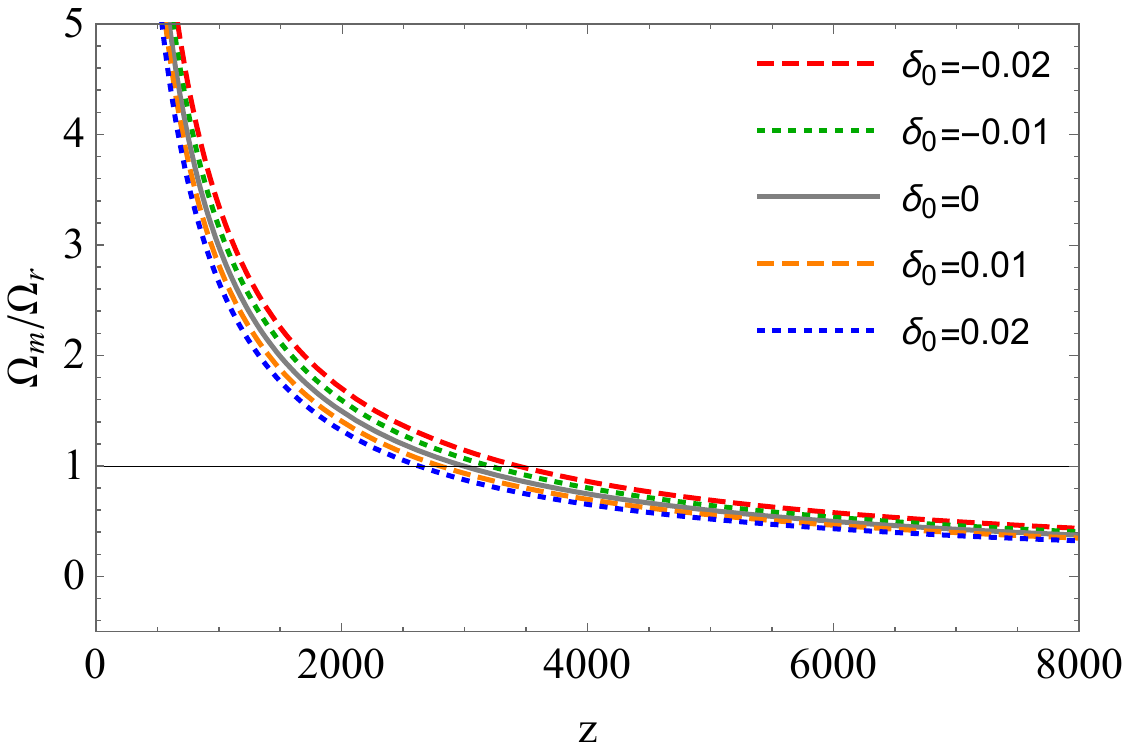}
    \caption{Redshift evolution of $\Omega_m/\Omega_r$ where $\Omega_m = \Omega_{\rm dm} +\Omega_b$, for different values of the interaction parameter $\delta_0$. The horizontal solid line corresponds to the matter radiation equality (i.e. $\Omega_m = \Omega_r$). In order to draw the curves we fix the other parameters  from the constraints obtained from  CMB+CC+DESI+PantheonPlus.  }
    \label{fig:matter-radiation}
\end{figure}
\begin{figure}
\includegraphics[width=0.48\textwidth]{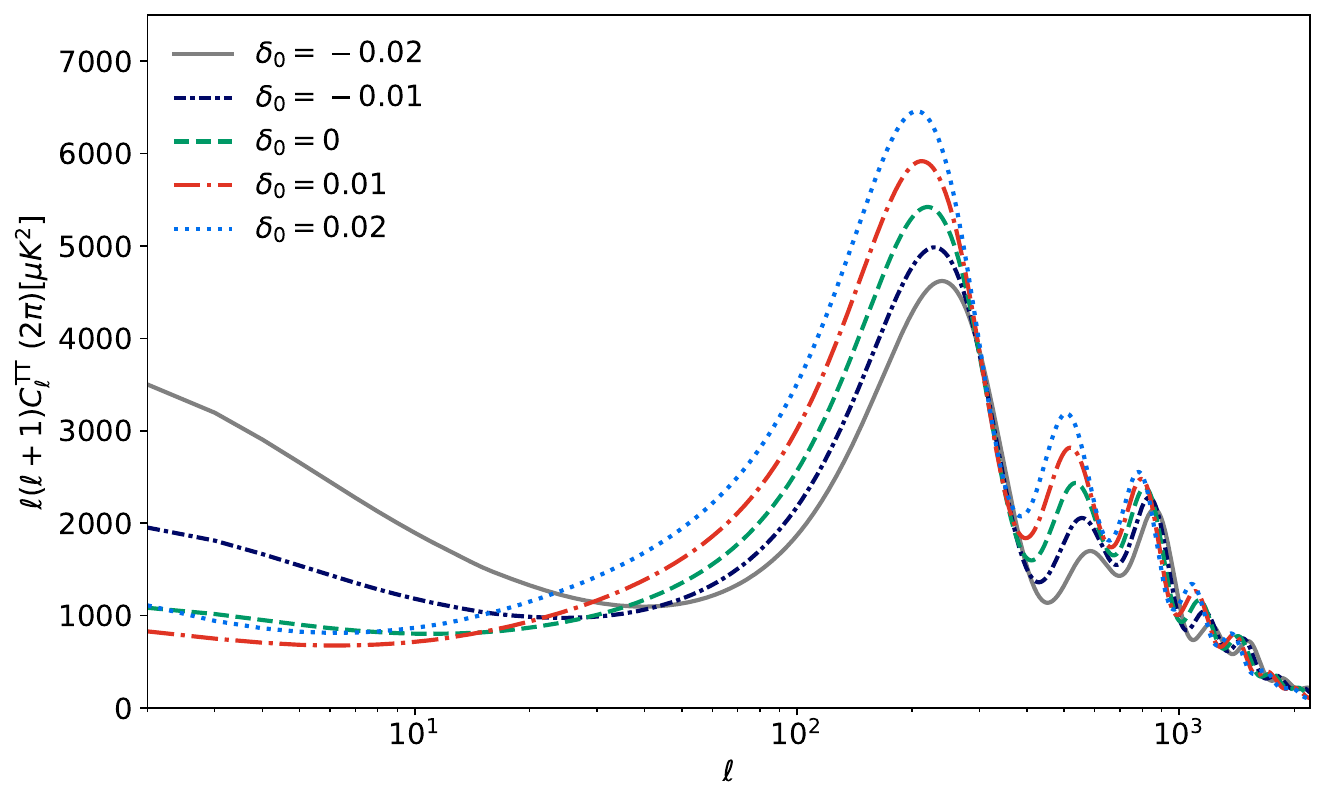}
\includegraphics[width=0.48\textwidth]{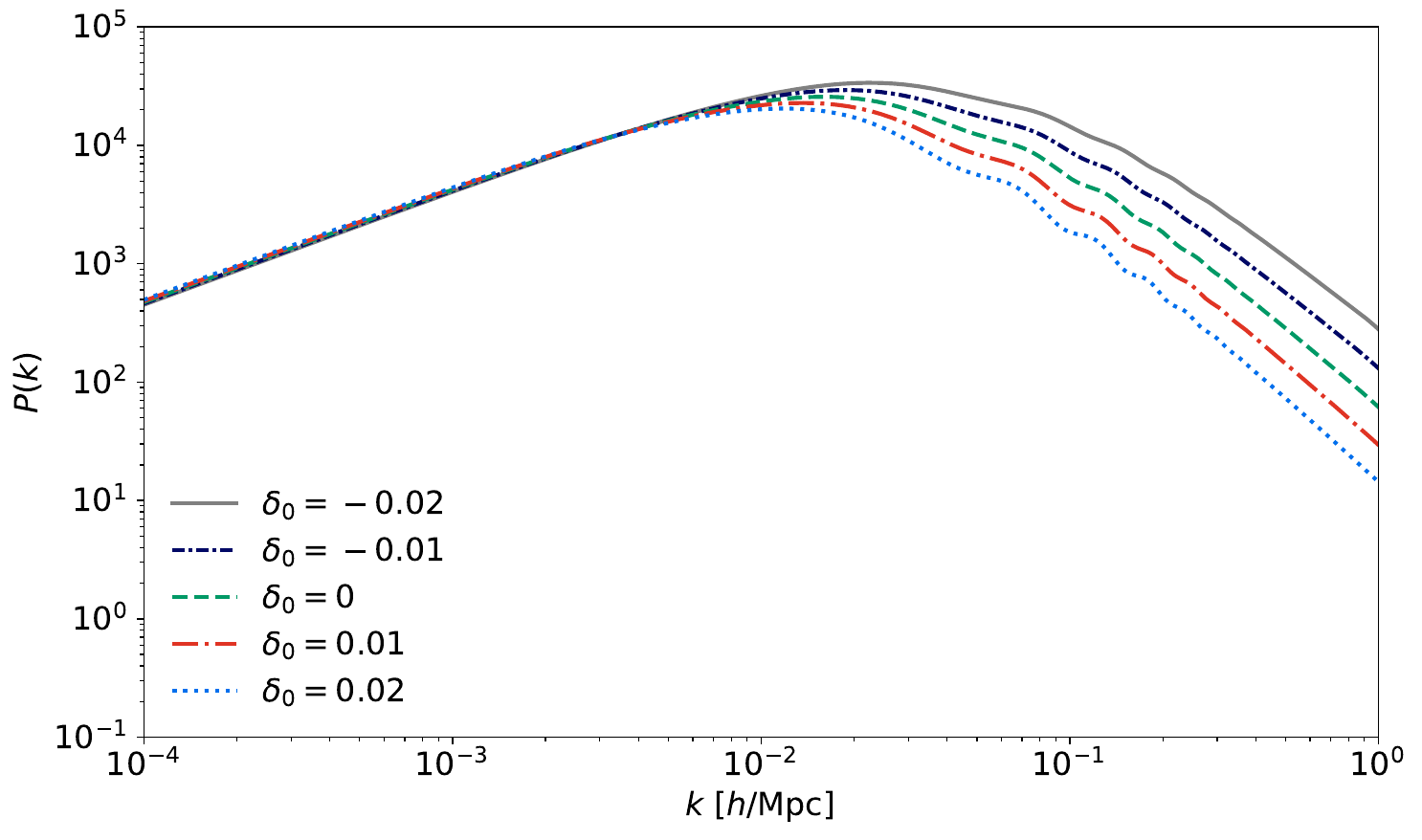}
    \caption{The CMB TT power spectrum (upper graph) 
    and the matter power spectrum (lower graph)
    for different values of $\delta_0$. In order to  draw the curves, we fix the 
    other parameters  from the constraints obtained from  CMB+CC+DESI+PantheonPlus. }  
    \label{fig:CMB-Matterpower}
\end{figure}

In this section  we proceed to  the detailed confrontation of the scenario of 
interacting dark energy with observational data.
In particular, we use the following data sets:
\begin{itemize}

\item Cosmic Microwave Background (CMB) measurements from Planck 2018 (PR3) release, incorporating the high-\textit{l} Temperature--Temperature (TT), Temperature--E-mode (TE), and E-mode--E-mode (EE) power spectra through the \textsc{Plik} likelihood, along with the PR3 low-\textit{l} TT and EE likelihoods~\cite{Planck:2018vyg, Planck:2018nkj, Planck:2019nip}.

\item Baryon Acoustic Oscillation (BAO) distance measurements
from ({\bf DESI}) DR2~\cite{DESI:2025zgx}.\footnote{\href{https://github.com/CobayaSampler/bao_data/tree/master/desi_bao_dr2}{https://github.com/CobayaSampler/desi\_bao\_dr2}} 

\item  Supernovae Type Ia (SNIa) from three well known astronomical surveys: (i)  {\bf PantheonPlus} from~\cite{Scolnic:2021amr, Brout:2022vxf};  
(ii) {\bf Union3} dataset consisting 2087 SNIa~\cite{Rubin:2023ovl}; (iii) five-year dataset of {\bf Dark Energy Survey} (labeled as {\bf DES-Y5}), covering the redshift range $0.025 < z <1.3$ \cite{DES:2024jxu}, along with 194 low-redshift SNeIa in the range $0.025 < z < 0.1$ ~\cite{Hicken:2009df, Hicken:2012zr, Krisciunas:2017yoe, Foley:2017zdq}.

\item Hubble parameter  measurements obtained from Cosmic Chronometers ({\bf CC}) approach, spanning the redshift range $0.179 < z < 1.965$ \cite{Moresco:2012by, Moresco:2015cya, Moresco:2016mzx}, incorporating the full covariance matrix, including non-diagonal correlations, and accounting for all systematic uncertainties.

    \end{itemize} 

 In order to constrain the interacting scenario, 
we conduct a comprehensive Markov Chain Monte Carlo (MCMC) investigation utilizing the publicly available cosmological tools, namely \textsc{CAMB}~\cite{Lewis:2002ah,Lewis:1999bs} and  \textsc{Cobaya}~\cite{Torrado:2020dgo}. The convergence of all the chains is carefully monitored through the Gelman-Rubin diagnostic test \cite{Gelman:1992zz}, confirming that the condition $R-1 < 0.02$. In this analysis, the model is characterized by seven independent parameters: one coupling parameter $\delta_0$, together with six baseline $\Lambda$CDM parameters -- specifically, the physical densities of dark matter ($\Omega_{\rm dm} h^2$) and baryons ($\Omega_{\rm b} h^2$),  optical depth due to reionization ($\tau$), the ratio of the sound horizon to the angular diameter distance at decoupling ($\theta_{\mathrm{MC}}$), the scalar amplitude of the primordial perturbations ($A_{\mathrm{s}}$), and the scalar spectral index ($n_{\mathrm{s}}$).  The uniform (flat) prior ranges assigned to these parameters during the MCMC sampling are detailed in Table~\ref{tab:priors}.

\begin{figure*}
    \centering
\includegraphics[width=\linewidth]{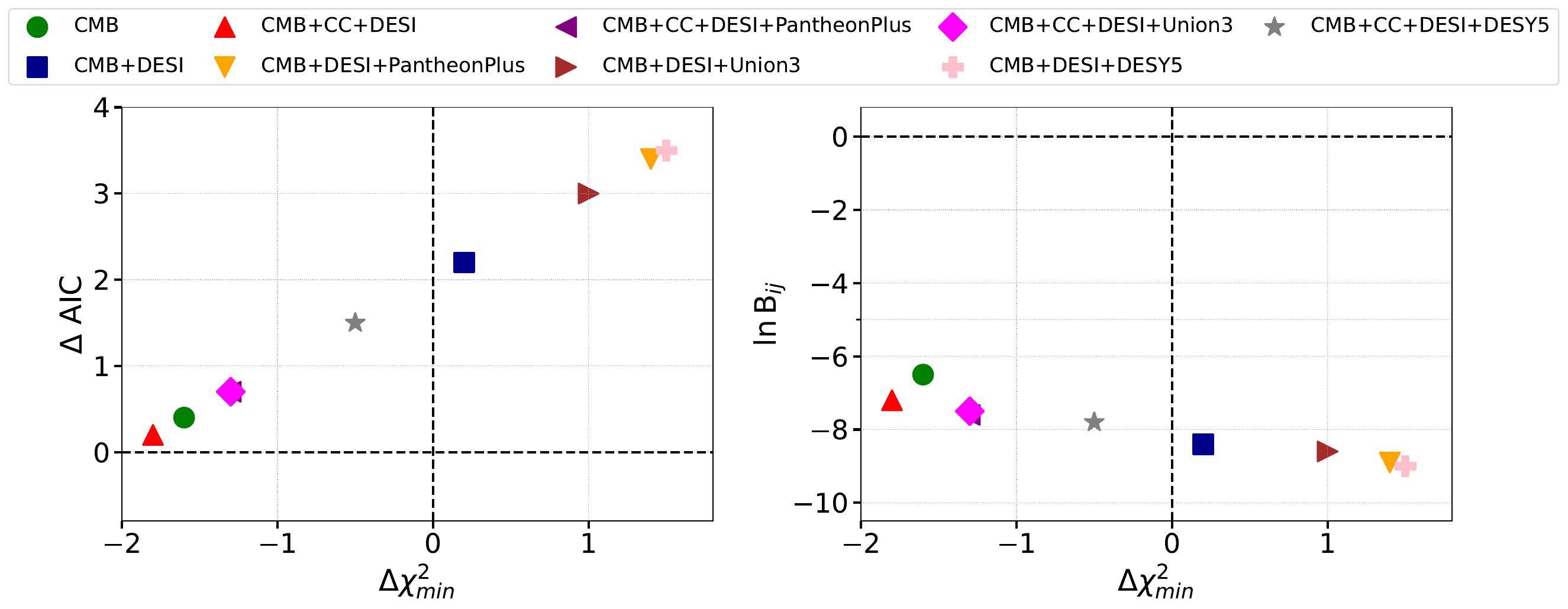}
    \caption{Graphical descriptions of various statistical measures ($\Delta$AIC and $\ln B_{ij}$) with respect to $\Delta \chi^2_{\rm min}$, across all the datasets.   }  
    \label{fig:IC}
\end{figure*}

The results are summarized in Tables~\ref{tab:IDE-noCC}, \ref{tab:IDE-withCC}  and \ref{tab:evidence} considering CMB alone and its combination with BAO from DESI DR2, three different compilations of SNIa data (PantheonPlus, Union3 and DESY5) and CC. Additionally,  in Figs. \ref{fig:contour1}, \ref{fig:contour2} we present the corresponding likelihood contours while in Fig.~\ref{fig:whisker}, we display the whisker plot  for the interaction parameter $\delta_0$,  Hubble parameter $H_{0}$ and matter density parameter at present $\Omega_{m0}$ ($=\Omega_{\rm dm,0}+\Omega_{b,0}$),  for the various datasets used in our analysis.  In Fig.~\ref{fig:CMB-Matterpower} we show the effects of the coupling parameter on the CMB TT spectra and matter power spectra and finally in Fig. \ref{fig:IC} we show a graphical description focusing on the preference of the model with respect to the standard $\Lambda$CDM scenario. 

In particular, in Table~\ref{tab:IDE-noCC} we present the constraints on the model parameters considering  CMB alone and   its combination with other datasets (excluding CC data), and in Table~\ref{tab:IDE-withCC} we have added CC with various combinations of the datasets,
in order to fully understand the extra constraining power, if any, coming from CC.

We begin with the constraints of Table~\ref{tab:IDE-noCC}. 
From CMB alone we see that a mild evidence of the interaction  parameter $\delta_0$  is found at slightly more than 68\% CL. Specifically, we find that  $\delta_0 > 0$ at more than 68\% CL, and thus  CMB alone suggests an energy flow from DE to DM. However, within 95\% CL we find that for $\delta_0$ the zero value
is allowed,  hence no strong evidence of interaction is found. When DESI is combined with CMB, evidence of interaction at more than 68\% CL is found, however, in this case, we find that the sign of $\delta_0$ is altered, namely $\delta _0< 0$ at more than 68\% CL and thus it indicates that the direction of energy flow is reversed (energy flows from DM to DE). 
Moreover, similarly to the CMB alone case, within 95\% CL  $\delta_0$ is allowed to acquire the zero
value,  which is in favor of a non-interacting model.
Finally, when three different compilations of the SNIa data are independently added to CMB+DESI, we see that our conclusions remain similar to CMB+DESI, namely  in all three cases (i.e. CMB+DESI+PantheonPlus, CMB+DESI+Union3 and CMB+DESI+DESY5) the interaction parameter $\delta_0$ remains non-null at more than 68\% CL and   an energy flow from DM to DE is indicated. 

We proceed  focusing on the constraints on $H_0$ and $\Omega_{m0}$ obtained from all the datasets,
summarized in Table~\ref{tab:IDE-noCC}. We note that for CMB alone, the model returns a very low value of $H_0$ ($= 64.06 \pm 1.79$ Km/s/Mpc at 68\% CL) and a high value of $\Omega_{m0}$ ($= 0.3623_{-0.0302}^{+0.0256}$ at 68\% CL). This can be easily understood by looking at the interaction flow. Since for CMB alone the energy flow takes place from DE to DM,  the DM energy density  increases,
which results to a higher value of $\Omega_{m0}$ and thus due to the geometrical degeneracy between $\Omega_{m0}$ and $H_0$ (see Fig. \ref{fig:contour1}) this scenario leads to a lower value of $H_0$. Consequently, this gives rise to a higher $S_8$ value of $0.866\, \pm \, 0.024 $ at 68 \% CL, which is somewhat elevated as compared to Planck 2018 estimate~\cite{Planck:2018vyg}. On the other hand, for the remaining cases in Table~\ref{tab:IDE-noCC}, the energy flow occurs from DM to DE (as $\delta_0 <0$), and thus we find a slightly lower value of $\Omega_{m0}$, which is compensated by the increase in the Hubble constant ranging between $68.42$ to $68.74$ Km/s/Mpc (note that Planck 2018 assuming $\Lambda$CDM in the background returns $H_0 =  67.27 \pm 0.60$ Km/s/Mpc at 68\% CL, Planck TT, TE, EE+lowE~\cite{Planck:2018vyg}). Furthermore, in all these cases, the corresponding $S_8$ values are almost similar to the Planck 2018 results~\cite{Planck:2018vyg} .

Let us consider the effects of the CC dataset when combined with CMB+DESI and CMB+DESI+SNIa (PantheonPlus, Union3, DESY5). In  Table~\ref{tab:IDE-withCC} we present the constraints on the model parameters. We see that for all the combined datasets  a mild evidence of interaction (at more 68\% CL) is found,
and  $\delta_0$ remains negative across all the combined datasets,   thus an energy flow from DM to DE is suggested. 
As a result, this interacting scenario leads to a slightly reduced value of $\Omega_{m0}$ and consequently to a
slightly high value of $H_0$ compared to the $\Lambda$CDM-based Planck 2018~\cite{Planck:2018vyg}. In Fig.~\ref{fig:whisker} we present a Whisker plot for $\delta_0$, $H_{0}$ and $\Omega_{m0}$,
considering all the datasets (without and with CC). 
For the interaction parameter $\delta_0$, two key observations emerge: first, the evidence of interaction at more than 68\% CL is pronounced irrespectively of all the datasets; secondly, we find that the inclusion of CC does not affect the interaction parameter. 
Furthermore, we find that the inclusion of CC does not affect the other model parameters. 
Additionally, the effect of the interaction flow on $H_0$ and $\Omega_{m0}$ is evident. A positive $\delta_0$ corresponds to a lower $H_{0}$ and a higher $\Omega_{m0}$, while a negative $\delta_0 $ shows the opposite trend, reinforcing  that the direction of energy exchange between dark sectors dictates the late-time acceleration behavior.

We now focus on the effects of the interaction
on the large scale structure of the universe,
through the  matter-radiation equality (see Fig. \ref{fig:matter-radiation}), the
CMB TT spectra and the matter power spectra (see Fig. \ref{fig:CMB-Matterpower}).  
We start with Fig. \ref{fig:matter-radiation},
showing the evolution of the ratio of matter  and radiation, 
$\Omega_m/\Omega_r$, for different (positive and negative) values of the interaction parameter 
$\delta_0$, including $\delta_0 =0$ which corresponds to the non-interacting $\Lambda$CDM scenario.  
We see that for $\delta_0 > 0$ the matter-radiation equality is slightly delayed,
since in this case  DM density decreases slowly compared to its standard evolution $a^{-3}$, and consequently the sound horizon increases, while for $\delta_0 <0$ the  matter-radiation equality occurs at an earlier epoch (and the sound horizon is therefore decreased). 
As a result, for $\delta_0 >0$  the height of the first peak of the CMB spectrum
increases comparing to the first peak obtained for $\delta_0 =0$, while  for $\delta_0 <0$ the height of the first CMB peak  decreases comparing to the case of $\delta_0 =0$ 
(see the upper graph of Fig. \ref{fig:CMB-Matterpower}). 
Additionally, the effect of the interaction is also observed in the low multipole region (large angular scales) of the CMB spectrum, where 
the integrated Sachs-Wolfe (ISW) effect is dominant. 
The interaction parameter $\delta_0$ 
affects the CMB spectrum through the ISW effect due to evolution in the gravitational potential. 
The effects of the interaction are equally visible in the matter power spectrum $P(k)$ (see the lower graph of Fig. \ref{fig:CMB-Matterpower}), 
where we notice that increasing positive values of $\delta_0$  lead to a suppression of power on small scales which is caused by the delayed matter–radiation equality. In contrast, negative    $\delta_0$ values advance the epoch of matter–radiation equality, enhancing the structure growth and amplifying the power on smaller scales.

\begin{table*}
        \centering
\resizebox{0.98 \textwidth}{!}{ %
    \huge  
\renewcommand{\arraystretch}{1.2} 
        \begin{tabular}{l @{\hspace{1cm}} c c c c c c c}
           \hline
              ~Dataset ~ &  ~Model~  &  ~$ \chi^2_{\rm min}$~ &  ~$\Delta \chi^2_{\rm min}$~ & ~AIC~ & ~$\Delta$AIC~  & $\ln {\rm B}_{ij}$  \\
            \hline
            \hline

                 ~\multirow{2}{*}{CMB}~& ~\texttt{Our-model}~ & ~$2778.7$~ & ~$-1.6$~ &  ~$2792.7$~ & ~$0.4$~ &  ~$-6.5$~ \\
                 & ~\texttt{$\Lambda$CDM}~ & ~$2780.3$~ & ~$...$~ & ~$2792.3$~ & ~$...$~ &  ~$...$~ \\
                
                 ~\multirow{2}{*}{CMB+DESI}~& ~\texttt{Our-model}~ & ~$2795.8$~ & ~$0.2$~ &  ~$2809.8$~ & ~$2.2$~ & ~$-8.4$~\\
                 & ~\texttt{$\Lambda$CDM}~ & ~$2795.6$~ & ~$...$~ & ~$2807.6$~ & ~$...$~ &  ~$...$~ \\

                ~\multirow{2}{*}{CMB+CC+DESI}~& ~\texttt{Our-model}~ & ~$2805.7$~ & ~$-1.8$~ & ~$2819.7$~ & ~$0.1$~ & ~$-7.2$~ \\
                & ~\texttt{$\Lambda$CDM}~ & ~$2807.5$~ & ~$...$~ & ~$2819.6$~ & ~$...$~  & ~$...$~ \\

                 ~\multirow{2}{*}{CMB+DESI+PantheonPlus}~ & ~\texttt{Our-model}~ & ~$4202.7$~ & ~$1.4$~ & ~$4216.7$~ & ~$3.4$~ & ~$-8.9$~\\
                 & ~\texttt{$\Lambda$CDM}~ & ~$4201.3$~ & ~$...$~ & ~$4213.3$~ & ~$...$~  & ~$...$~\\

                 ~\multirow{2}{*}{CMB+CC+DESI+PantheonPlus}~& ~\texttt{Our-model}~ & ~$4212.2$~ & ~$-1.3$~ & ~$4226.2$~ & ~$0.7$~  &  ~$-7.6$~ \\
                 & ~\texttt{$\Lambda$CDM}~ & ~$4213.5$~ & ~$...$~ & ~$4225.5$~ & ~$...$~  &  ~$...$ \\

                ~\multirow{2}{*}{CMB+DESI+Union3}~ & ~\texttt{Our-model}~ & ~$2825.1$~ & ~$1.0$~ & ~$2839.1$~& ~$3.0$~ & ~$-8.6$~  \\
                & ~\texttt{$\Lambda$CDM}~ & ~$2824.1$~ & ~$...$~ & ~$2836.1$~& ~$...$~   & ~$...$~ \\

                ~\multirow{2}{*}{CMB+CC+DESI+Union3}~ & ~\texttt{Our-model}~ & $2834.7$ & ~$-1.3$~ & ~$2848.7$~ &~$0.7$~ & ~$-7.5$~\\
                & ~\texttt{$\Lambda$CDM}~ & $2836.0$ & ~$...$~ & ~$2848.0$~ &~$...$~     & ~$...$~ \\

                ~\multirow{2}{*}{CMB+DESI+DESY5}~ & ~\texttt{Our-model}~  & $4446.6$ & ~$1.5$~ & ~$4460.6$~ & ~$3.5$~  &  ~$-9.0$~ \\
                & ~\texttt{$\Lambda$CDM}~ & $4445.1$ & ~$...$~ & ~$4457.1$~ & ~$...$~    & ~$...$~ \\

                ~\multirow{2}{*}{CMB+CC+DESI+DESY5}~ & ~\texttt{Our-model}~  & $4456.5$ & ~$-0.5$~ & ~$4470.5$~ & ~$1.5$~  & ~$-7.8$~ \\
                & ~\texttt{$\Lambda$CDM}~ & $4457.0$ & ~$...$~ & ~$4469.0$~ & ~$...$~    & ~$...$~ \\

            \hline
            \hline
        \end{tabular}}
         \caption{Summary of $\Delta{\chi}^2_{\rm min}$,  $\Delta$AIC, $\ln {\rm B}_{ij}$ for several observational datasets.  Negative values of $\Delta\chi^2_{\rm min}$ and $\Delta$AIC, as well as positive values of $\ln {\rm B}_{ij}$, indicate a statistical preference for the interacting dark energy scenario over the $\Lambda$CDM reference model.  }
    
        \label{tab:evidence}
    \end{table*}

We close this section by performing a statistical comparison of the interacting dark energy scenario with the standard non-interacting case, namely the $\Lambda$CDM paradigm, applying    
 the Akaike Information Criterion (AIC)~\cite{1100705},\footnote{The AIC is defined as: $\text{AIC} \equiv -2 \ln \mathcal{L}_{\text{max}} + 2k$,  where $\mathcal{L}_{\text{max}}$ represent the maximum likelihood of the model for the considered dataset and $k$  represents the number of free parameters for considered model.}  and the Bayesian evidence analysis~\cite{Trotta:2008qt}.  To quantify the relative preference between the interacting model and $\Lambda$CDM, we evaluate the differences $\Delta \chi^2_{\rm min}$ (where $\Delta \chi^2_{\rm min} \equiv$  $\chi^2_{\rm min}$ (Model) $-$  $\chi^2_{\rm min}$ ($\Lambda$CDM)) and $\Delta \text{AIC}$ (where $\Delta$AIC $\equiv$ AIC (Model) $-$ AIC ($\Lambda$CDM). If $\Delta \text{AIC} < 0$ the model is strongly favored; values in the range $0 < \Delta \text{AIC} < 2$ indicate that the models are statistically indistinguishable; $4 < \Delta \text{AIC} < 7$ reflects considerably less support for the model, while $\Delta \text{AIC} > 10$ suggests that the model is strongly disfavored compared to the reference model \cite{burnham2003model}. 

 On the other hand, in order to calculate 
the Bayesian evidence we use the \textsc{MCEvidence} package~\cite{Heavens:2017hkr,Heavens:2017afc}. The Bayesian inference for a given model $\mathcal{M}_i$ defined by its parameter vector $\Theta$, is expressed as $B_i = \int \mathcal{L}(D|\Theta,\mathcal{M}_i)\,\pi(\Theta|\mathcal{M}_i)\,d\Theta$, where $\mathcal{L}$ denotes the likelihood of the data $D$ and $\pi$ represents the prior probability of the parameters. To compare the competing models, one calculates
the Bayes factor $B_{ij} = B_i / B_j$ ($j$ refers to the reference model and here we consider our reference model to be the $\Lambda$CDM), or equivalently the difference in their logarithmic evidences, $\ln B_{ij} = \ln B_i - \ln B_j$, with $\ln B_{ij} > 0$ indicating the preference of $M_i$ over $M_j$ (i.e. $\Lambda$CDM). The strength of model preference is quantified using the modified Jeffreys’ scale~\cite{Kass:1995loi}, where $\ln B_{ij} \lesssim 1$ indicates an inconclusive result, $1 \lesssim \ln B_{ij} \lesssim 2.5$ corresponds to weak support, $2.5 \lesssim \ln B_{ij} \lesssim 5$ denotes a moderate preference, and $\ln B_{ij} > 5$ signifies strong evidence. 

Finally, the interacting model is statistically preferred over $\Lambda$CDM when $\Delta \chi^2_{\rm min} < 0$, $\Delta \text{AIC} < 0$, and $\ln B_{ij} > 0$.

Building upon the statistical framework for model comparison, the results obtained from various dataset combinations are summarized in  Table~\ref{tab:evidence}. In terms of the $\Delta \chi^2$ estimation, the base CMB dataset shows a mild preference for our model. The inclusion of DESI data slightly shifts the trend toward the $\Lambda$CDM scenario, 
though still within the mild regime. For the combined CMB+CC+DESI dataset, the preference again leans in favor of the interacting model. In contrast, the CMB+DESI+PantheonPlus dataset mildly supports the $\Lambda$CDM model, whereas the inclusion of CC data in this combination (i.e. for CMB+CC+DESI+PantheonPlus) shifts the indication back toward the interacting scenario, remaining within the mild significance regime. Now in view of $\Delta{\rm AIC}$, an indication for the preference of $\Lambda$CDM over the interacting scenario is noticed (as $\Delta{\rm AIC} >0$ for all the datasets); however, for some of the datasets, distinguishing between the models becomes difficult (since $0<\Delta{\rm AIC} < 1$).  While on the contrary, according to  $\ln B_{ij}$ estimates, $\Lambda$CDM remains preferred over the interaction model. To aid interpretation, Fig.~\ref{fig:IC} shows the variations of $\Delta{\rm AIC}$,  and $\ln B_{ij}$ against $\Delta \chi^2_{\rm min}$, offering a complementary visualization of the statistical consistency across different dataset combinations. 

In summary, the above observational confrontation and model comparison reveals a mixed picture   on the overall selection between the interaction model and the $\Lambda$CDM paradigm.

\section{Summary and Conclusions}
\label{sec-summary}

In this letter we considered a general interacting scenario between DM and DE,  quantified  through the deviation 
from the standard scaling of the DM energy density, $\rho_{\rm dm} \propto a^{-3 + \delta(a)}$ where $\delta (a) = {\it const}$. Considering the most basic interacting scenario in which DM is pressureless and DE is the vacuum energy density, we constrained the scenario employing the latest astronomical probes including CMB from Planck 2018, BAO from DESI DR2, SNIa datasets (PantheonPlus, Union3, DESY5) and Hubble parameter measurements from CC. According to our results, a mild interaction in the dark sector is suggested at more than $1\sigma$, while within $2\sigma$, $\Lambda$CDM is recovered.  
However, comparison  with $\Lambda$CDM scenario through $\Delta\chi^2_{\rm min}$,  AIC and Bayesian analysis, reveals a mixed picture, where we find that according to $\Delta \chi^2_{\rm min}$, most of the datasets prefer the interacting scenario over the $\Lambda$CDM, while other statistical measures go in favor of the $\Lambda$CDM.

Our results may reveal evidence of interaction in the dark sector.  The evidence of interaction may increase once  $\delta (a)$ is considered to be  time-varying, which is the most general case, or if a different interaction function is assumed \cite{Shah:2025ayl,Silva:2025hxw}, or if DE is considered to be more general than  simple vacuum energy. We mention that since  the interaction may  lead to an effective DM equation-of-state parameter deviating from zero, which may lead to the possibility of a non-cold DM \cite{Pan:2022qrr}, the fact that  DESI DR2 favors dynamical DE~\cite{DESI:2025zgx} perhaps leads to  an evidence of interaction together with a non-cold DM.  The question whether the dark sectors interact mutually remains open.

\section*{Acknowledgments}
We thank an anonymous  referee for   fruitful comments that assisted us improve the quality of the manuscript. We also thank Eleonora Di Valentino for useful correspondence. 
SP acknowledges the support of the Department of Science and Technology (DST), Govt. of
India under the Scheme ``Fund for Improvement of S\&T Infrastructure (FIST)'' 
(File No. SR/FST/MS-I/2019/41). ENS acknowledges the contribution of the LISA 
CosWG, and of   COST 
Actions  CA18108  ``Quantum Gravity Phenomenology in the multi-messenger 
approach''  and  CA21136 ``Addressing observational tensions in cosmology with 
systematics and fundamental physics (CosmoVerse)''.  W. Y has been supported by the National Natural Science Foundation of China under Grant Nos. 12547110 and 12175096.  
We acknowledge the computing facility of the ICT Department of Presidency University, Kolkata.

\bibliography{references}

\end{document}